# Energy Storage System Design for a Power Buffer System to Provide Load Ride-through

X.Y. Wang, *Member,* D. M. Vilathgamuwa, *Senior Member,* and S. S. Choi, *Senior Member*

*Abstract*-- The design of a power buffer to mitigate the negative impact of constant power loads on voltage stability as well as enhancing ride-through capability for the loads during upstream voltage disturbances is examined. The power buffer adjusts its front-end converter control so that the buffer-load combination would appear as a constant impedance load to the upstream supply system when depressed voltage occurs. A battery energy-storage back-up source within the buffer is activated to maintain the load power demand. It is shown that the buffer performance is affected by the battery state of discharge and discharge current. Analytical expressions are also derived to relate the buffer-load ride-through capability with the battery state-of-discharge. The most onerous buffer-battery condition under which the load-ride through can be achieved has been identified.

*Index Terms*-- Constant power load, lead-acid battery, power buffer, power quality, voltage stability.

## I. INTRODUCTION

With the increasing wide-spread use of power electronics-based loads in electrical supply systems, important contributing factor of these loads to system stability has been recognized for some time [1]. When an upstream fault occurs and causes depressed voltage or sag, the constant power loads will draw a larger current from upstream source in order to maintain the demand. This in turn causes a larger voltage drop across the source impedance and to a further decrease in the load terminal voltage. Depending on the severity of the sag and the load demand level, the situation could develop into voltage instability or even a system collapse.

To mitigate the aforementioned voltage stability problem, active dynamic power buffer concept has been proposed to decouple the load dynamics from the upstream system during fault. In [2, 3], the authors proposed a power buffer scheme in which a capacitor is used as the form of energy-storage media to balance the mismatch in power between that supplied from upstream system and the load demand during the disturbance interval. The authors assume that the capacitor is of sufficient capacity to meet the power shortfall. Under normal conditions, the buffer scheme in [2, 3] is controlled to maintain unity power factor (UPF) at the point of common coupling (PCC). When voltage sag occurs, however, the buffer in [2, 3] adjusts the front-end converter so that the input impedance of the buffer (observed from the source side) is controlled to assume the same value as that before sag. Hence the input current would decrease proportionally as the load terminal voltage reduces. In this way, voltage drop across the source impedance would also be reduced and the voltage instability problem could be mitigated. The buffer scheme is extended in [4] to deal with unbalanced faults but with the input impedance during disturbance controlled so that the input current is balance and stays within set limit.

In all the above works, it is clear that if the severity of the disturbance is such that the upstream supply system is unable to meet the load demand, the capacitor energy storage device within the buffer must provide the shortfall. Unfortunately, all the analysis in [2-4] has not specifically addressed the design of the energy storage system. The intent of this paper is to fill this gap. In the proposed scheme, a battery energy storage system is used as the back-up power source. In Section II, the power buffer and its operation scheme are presented. Load ride-through capability, as it relates to the battery state of discharge, is analyzed in Section III. Simulation results are given in Section IV to show the effectiveness of the proposed scheme.

## II. POWER BUFFER OPERATIONAL PRINCIPLE

There are several topologies proposed for the power buffer. The specific version shown in Fig. 1 is comprised of a three-phase boost converter, controlled through the PWM switching scheme and a series RL filter. The filter controls the level of harmonic generated by the converter. Usually the filter resistance is small compared to the inductance and can be neglected. Such a converter topology would allow it to operate in either rectifying or regenerating modes. It could be controlled to draw almost sinusoidal input current and with a near unity power factor at its terminals. In this system, $V_g$ represents the upstream source-side voltage at the point-of-common-coupling (PCC) of the buffer-load combination. Also, it is assumed that the filter is effective and that of the impedances shown in Fig. 1, only their components at the mains power frequency ($\omega$) need to be considered.

Under normal condition, the PWM switching scheme is controlled to maintain the dc-link voltage ($V_{dc}$) at desirable level and to meet the constant power needed by the load. The control scheme of the buffer system described in Figure 1 is

---

X. Y. Wang is with Sustainable Energy Technologies Department of Brookhaven National Laboratory, Upton, NY 11961, USA. D. M. Vilathgamuwa and S. S. Choi are with the School of Electrical and Electronic Engineering, Nanyang Technological University, Singapore 639798, Singapore (emails: xywang@bnl.gov, emahinda@ntu.edu.sg and esschoi@ntu.edu.sg).

given in [4] where controllers have been designed to guarantee system stability and to keep $V_{dc}$ at nominal level.

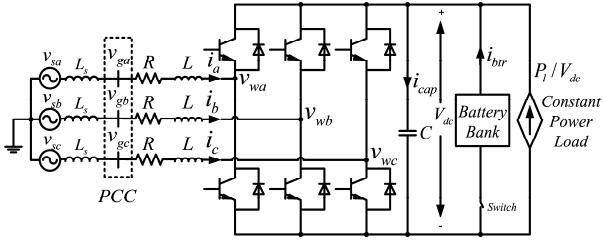

Fig. 1 A schematic diagram of a power buffer-connected load scheme.

When voltage sag occurs, the power buffer is designed such that the buffer-load combination is switched from the constant power mode to one of constant impedance. Under the latter mode, the buffer-load combination could be represented by a parallel connection of $R_{in}$ and $X_{in}$, as shown in Fig. 2. In the figure, $\vec{V}_s$ is the upstream voltage and $X_s$ represents the source impedance where $X_s = j\omega L_s$.

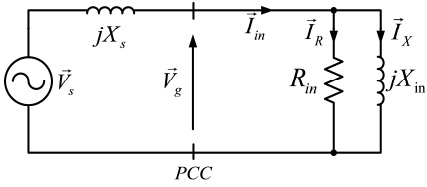

Fig. 2 Equivalent circuit of power buffer-load combination in parallel representation.

In Fig. 2, $R_{in}$ and $X_{in}$ are determined through the following equations,

$$R_{in} = V_{g0}^2 / P_l \quad (1)$$
$$X_{in} = V_{g0}^2 / Q_l \quad (2)$$

where $P_l$ and $Q_l$ are the constant real and reactive power load components respectively. $V_{g0}$ is the pre-sag voltage at the PCC. Note that under UPF operation, $X_{in}$ value is very much larger than $R_{in}$. also under steady-state, $R_{in}$ and $X_{in}$ can be readily determined using (1) and (2) because $V_{g0}$, $P_l$ and $Q_l$ are measurable. The determined values would be the equivalent impedance of the load-buffer combination prior to voltage sag.

With the RL filter, the filter impedance $jX$ forms part of the input impedance $R_{in}$ and $X_{in}$. Representing the buffer-load impedance as series connection of $R_1$ and $X_1$, $R_1$ and $X_1$ can be expressed in terms of $R_{in}$ and $X_{in}$ using the following expressions [4]:

$$\left. \begin{array}{l} R_1 = R_{in} X_{in}^2 / (R_{in}^2 + X_{in}^2) \\ X_1 = R_{in}^2 X_{in} / (R_{in}^2 + X_{in}^2) - X \end{array} \right\} \quad (3)$$

The buffer terminal voltage phasor $\vec{V}_w$ is

$$\vec{V}_w = \vec{V}_g (R_1 + jX_1)/[R_1 + j(X + X_1)]$$

Taking $\vec{V}_g$ as the reference phasor and after some manipulation, the d, q components of $\vec{V}_w$ can be derived:

$$\left. \begin{array}{l} V_{wd} = V_{gd}[R_1^2 + X_1(X + X_1)]/[R_1^2 + (X + X_1)^2] \\ V_{wq} = -V_{gd} R_1 X /[R_1^2 + (X + X_1)^2] \end{array} \right\} \quad (4)$$

Real and reactive power flows into the power buffer can be regulated through the control on $\vec{V}_w$ in the following way. Under fault condition, one attempts to maintain the input impedance at the load-buffer terminal constant at the known pre-sag value $R_{in} + jX_{in}$. This is achieved through controlling the buffer PWM to produce $\vec{V}_w$ given in (4). $R_1$ and $X_1$ are readily obtained through (3) and as $\vec{V}_g$ can be measured on-line, hence $\vec{V}_w$ can be generated. In this way, the power drawn by the buffer is governed by $R_{in}$ and $X_{in}$ and is not determined by the power absorbed by the load. In this way, the buffer-load combination is viewed as constant impedance from the source.

The approach described differs from that in [4]. Whereas in controlling power flow from the upstream source into the PCC, the approach in [4] requires the input current be controlled to within pre-set limit, the present approach achieves the power flow control by ensuring the input impedance of the buffer-load combination is exactly the same as that pre-fault. In the latter approach, the power flow into the PCC will be proportional to the square of the PCC voltage magnitude, $V_g$. Therefore, the input power drawn by the buffer-load connection will decrease during the sag event. Any mismatch in power between that from the PCC and the load demand can be compensated by the energy storage system connected across the dc-link, thus creating a favorable situation in terms of system stability.

On the control system based on the above concept to cater for both balanced and unbalanced sags, negative phase sequence controller is design in the same manner as for the positive phase sequence system. The method is similar to that given in [4].

### III. A NEW POWER BUFFER SCHEME

From Fig. 2, the phasor diagram depicted in Fig. 3(a) can be obtained. When an upstream fault occurs and since the input impedance of the buffer-load combination remains constant at the pre-sag value, $I_{in}$ would decrease in proportion with the voltage sag depth. From Fig. 3(a) and for a given sag, maximum real power drawn from $V_s$ occurs when the power flow from source at the PCC is at unity power factor (UPF), i.e., $\vec{I}_{in}$ is in phase with $\vec{V}_g$. The phasor diagram depicting such maximum power transfer condition is shown in Fig. 3(b) and the corresponding power transfer level is

$$P_{max} = V_g^2 / R_{in} \quad (5)$$

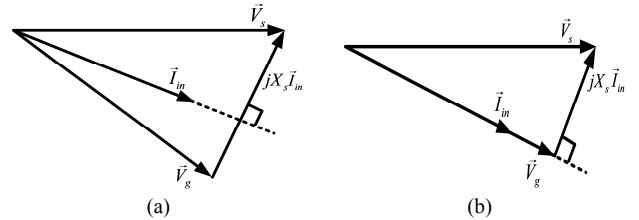

(a)                      (b)

Fig. 3 Phasor diagram describing power buffer at (a) non-unity power factor and (b) unity power factor condition at the PCC.

Hence, in the proposed scheme, once a sag is detected, i.e. $V_g < V_{g,min}$, where $V_{g,min}$ is typically 0.95 p.u. [5], the power buffer is switched from constant power mode to one of constant impedance mode.

Next the focus is to examine the requirement of the energy storage system when the buffer operates under the constant impedance mode.

*A. Circuit Model of Buffer with Battery Storage*

Unlike [2, 3] where capacitor has been proposed as the media of energy storage, battery has been considered for incorporation in the buffer scheme in this investigation. With its competitive price, lead-acid battery has become one widely used energy storage device in power systems. Moreover, battery energy storage system possesses the merit of higher specific energy capacity when compared to capacitor [6]. Hence it is likely to be suitable for the present high power application.

The proposed buffer scheme will now be described. Several lead-acid battery models have been reported in the literature [7-11]. One most commonly used model is the Thevenin equivalent representation shown in Fig. 4. In the model, $E$ is the no-load battery voltage and $R_s$, the internal resistance. $R_p$ represents the non-linear contact resistance of the plate to electrolyte and $C_p$, the capacitance of the parallel plates of the electrolyte and electrode. In practice, these battery parameters are dependent of the state of discharge ($f$) of the battery, the level of the discharge current and the battery operating temperature. In the present analysis, however, since it is expected that the battery is to function over short durations, one may assume $E$, $R_s$, $R_p$ and $C_p$ are constant over the operational period. The battery model would then be suitable for use in analyzing the behavior of the buffer.

From load side of Fig. 1 and Fig. 4, one can obtain the following equations based on basic circuit laws,

$$\begin{aligned} i_{cp} &= C_p dv_{cp}/dt, \quad i_{cap} = C dv_{dc}/dt, \quad i_{btr} = i_{cp} + v_{cp}/R_p, \\ i_{btr} &= (E - v_{dc} - v_{cp})/R_s, \quad i_{btr} - i_{cap} = \Delta p/v_{dc} \end{aligned} \quad (6)$$

where $v_{dc}$, $v_{cp}$ are the voltages across the dc-link and that across the battery capacitance $C_p$ respectively. $i_{btr}$ is the current in the battery branch, $i_{cp}$ is that in the capacitor sub-branch and $i_{cap}$ is that in the dc-link capacitor. Let $\Delta p(t)$ be the mismatch power between the power buffer and the constant power load. Under the UPF operating condition at the buffer terminals,

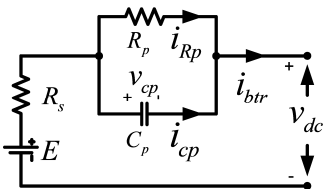

Fig. 4 Circuit model of lead-acid battery bank.

$$\Delta p(t) = P_l - P_{max} \quad (7)$$

$P_l$ represents the load power demand. $P_{max}$ is the maximum power defined in (5) and is proportional to $V_g^2$.

After some algebraic manipulations of (5)-(7), one obtains the following equations to describe the dynamic behavior of the power buffer over the transient period,

$$\left.\begin{aligned} 0.5C\, dv_{dc}^2/dt &= -\Delta p - v_{dc}v_{cp}/R_s + Ev_{dc}/R_s - v_{dc}^2/R_s \\ C_p\, dv_{cp}/dt &= E/R_s - (R_s + R_p)v_{cp}/(R_s R_p) - v_{dc}/R_s \end{aligned}\right\} \quad (8)$$

*B. Steady-State Analysis of Power Buffer Operation*

From (8), the equations describing the steady state of the buffer are

$$\left.\begin{aligned} \Delta P + V_{dc,ss}V_{cp,ss}/R_s + V_{dc,ss}^2/R_s - EV_{dc,ss}/R_s &= 0 \\ E/R_s - (R_s + R_p)V_{cp,ss}/(R_s R_p) - V_{dc,ss}/R_s &= 0 \end{aligned}\right\} \quad (9)$$

where $\Delta P$, $V_{dc,ss}$ and $V_{cp,ss}$ represent steady state values of $\Delta p$, $v_{dc}$ and $v_{cp}$ of (8) respectively.

From (9), the relationship between the steady state voltages ($V_{cp,ss}$) across $C_p$, and the mismatch power $\Delta P$ in terms of the dc-link voltage ($V_{dc,ss}$) and battery parameters are

$$V_{cp,ss} = R_p(E - V_{dc,ss})/(R_s + R_p) \quad (10)$$
$$\Delta P = V_{dc,ss}(E - V_{dc,ss})/(R_s + R_p) \quad (11)$$

In carrying out this analysis, it is assumed that the transient process following the switching in of the battery bank has ceased. This means that only the battery resistance needs to be considered. The universal battery model given in [10] is then applicable:

$$E = E_0 - Kf, \qquad R_b = R_0 - K_R f \quad (12)$$

In this model, $E_0$ and $R_0$ are the open circuit terminal battery voltage and the total equivalent battery resistance at full charge respectively. $R_b$ is the total internal resistance which depends on the battery state of discharge (SOD) or $f$. $K$ and $K_R$ are experimentally obtained constants. When the battery is fully charged, $f = 0$. $f > 0$ indicates that the battery is in the discharged state. Conversely, $f < 0$ shows the battery is over-charged.

Substituting (12) into (11), the relationship between $V_{dc,ss}$, $f$ and $\Delta P$ can be obtained.

$$V_{dc,ss} = 0.5[E_0 - Kf + \sqrt{(E_0 - Kf)^2 - 4(R_0 - K_R f)\Delta P}] \quad (13)$$

Equation (13) allows one to determine $V_{dc,ss}$ for given battery $f$ and $\Delta P$.

Furthermore, by substituting (5) and (7) into (13), one obtains

$$V_{dc,ss} = 0.5[E_0 - Kf + \sqrt{(E_0 - Kf)^2 - 4(R_0 - K_R f)(P_l - V_g^2/R_{in})}] \quad (14)$$

Equation (14) relates the variations in $V_{dc,ss}$ with the voltage magnitude $V_g$, the battery discharge state $f$ and the constant impedance ($R_{in}$) considered in Section II and is given by (1). Typical curves describing (14) are shown on Fig. 5. In this figure, it is seen that for a given sag voltage ($V_g$), $V_{dc,ss}$ increases with the decrease of $f$ (i.e. battery is more fully charged). This means that the battery with a lower $f$ will have a higher internal EMF, lower internal resistance and can therefore supply more power than a battery with higher $f$, for given $E_0$ and $R_0$.

In practice, to guarantee satisfactory operation of the buffer, it is also necessary to ensure that $V_{dc,ss}$ should be kept within a range (say) $\pm 10\%$ around its nominal value [12]. These limits are shown on Fig. 5 as curves AB and CD. These two boundaries define the operating range of the buffer for different battery SOD. Hence the feasible operating region of the buffer-load combination is within the area ABCD. One could use the information given in this figure to readily estimate the most severe sag or swell the buffer would be able to provide load ride-through under steady-state conditions. For example, if the battery SOD $f = 0.4$, from Fig. 5, one notes that any voltage disturbance which causes $V_g$ to lie within the range 0.82 p.u. (sag) to well beyond 1.4 p.u. (swell) can be readily dealt with by the buffer, and the load demand can be met with $V_{dc,ss}$ within the acceptable limit. As the battery SOD can be monitored, a possible implementation scheme would entail the on-line measurement of $f$, and subsequently use it to predict the load ride-through level. Since the buffer is used to provide the short-fall in power during the load ride-through, $f$ will increase over time. The ride-through capacity of the buffer would be updated continuously so that one could determine when the battery bank needs to be re-charged, in order to meet some pre-specified load ride-through criteria.

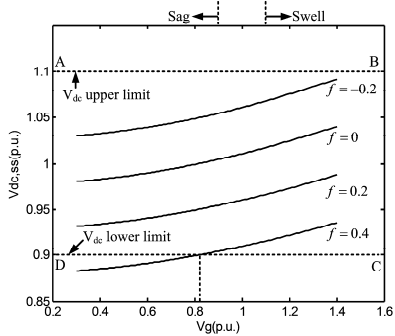

Fig. 5 Steady-state DC-link voltage ($V_{dc,ss}$) as function of battery state of discharge ($f$) and sag/swell level.

In fact, one should also consider the inevitable delay from the instance of the detection of the voltage disturbance until the sag is confirmed. Based on the ITI/CBEMA curve [13], typical satisfactory operation of sensitive loads can still be maintained if the delay is less than 2 ms and the dc-link voltage is quickly restored soon after. Over this initial sag interval of 2 ms, the load-buffer will be allowed to operate under constant power mode. This is because voltage instability is unlikely to be very much affected over this relatively short interval. Once the sag event is confirmed and its severity identified, the buffer-load combination is then switched into constant impedance mode.

While the above is concerned with voltage sag incident, similar analysis can be carried out for voltage swell, i.e. when $V_g > 1.1$ p.u. (say).

C. Transient Performance of Power Buffer

As mentioned in the last sub-section, one expects the variations of dc-link voltage ($V_{dc}$) are controlled to within a narrow range of its nominal value. Consequently, it is reasonable to analyze the transient performance of (8) in its small-signal form as,

$$\left.\begin{array}{l} d\delta v_{dc}/dt = -(2V_{dc,ss} + V_{cp,ss} - E)\delta v_{dc}/(R_s C V_{dc,ss}) \\ \quad - \delta v_{cp}/(R_s C) - \delta \Delta p/(C V_{dc,ss}) \\ d\delta v_{cp}/dt = -\delta v_{dc}/(R_s C_p) - (R_s + R_p)\delta v_{cp}/(C_p R_s R_p) \end{array}\right\} \quad (15)$$

Stability and transient performance of the system following the battery switch-in may now be analyzed. From (15), one can obtain the transfer function

$$\delta v_{dc}/\delta \Delta p = -(s+\alpha)/[CV_{dc,ss}(s^2 + (\alpha-\beta)s - \gamma - \alpha\beta)] \quad (16)$$

where,

$$\alpha = \frac{R_s + R_p}{R_s R_p C_p}, \quad \beta = \frac{E - 2V_{dc,ss} - V_{cp,ss}}{R_s C V_{dc,ss}}, \quad \gamma = \frac{1}{R_s^2 C C_p} \quad (17)$$

From the characteristic equation of (16) and using Hurwitz-Routh criterion, stability of the buffer-battery system would be guaranteed if and only if

$$\alpha - \beta > 0 \quad (18)$$
$$-\alpha\beta - \gamma > 0 \quad (19)$$

From (18), one obtains the necessary condition for stability as

$$V_{dc} > R_s R_p C_p E/[(R_s + R_p)^2 C + R_p C_p (2R_s + R_p)] \quad (20)$$

Substituting (10) into (19), $V_{dc,ss}$ must also satisfy the condition

$$V_{dc,ss} > 0.5E \quad (21)$$

Upon closer examination it is noted that the following in-equality relationship is always valid,

$$(R_s + R_p)^2 C + R_p C_p (2R_s + R_p) > 2R_s R_p C_p \quad (22)$$

Hence, one can conclude once (21) is satisfied, (20) would be satisfied automatically.

Under battery discharge mode, $\Delta P > 0$. From (11), it can be determined that indeed, $V_{dc,ss} > 0.5E$. Hence, (21) is always satisfied. Thus the buffer system is always stable and $v_{dc}$ will settle to a constant value following a step change $\Delta p$.

Furthermore, one can also examine the characteristic equation of (16) to obtain the damping ratio $\zeta$ where

$$\zeta = 0.5(\alpha-\beta)/\sqrt{-\alpha\beta - \gamma}$$

Substituting (17) into the last expression and since $E \approx V_{dc,ss}$, $\zeta$ can be expressed in terms of $R_s$, $R_p$ and $C_p$, viz

$$\zeta = 0.5(R_s C + R_p C + R_p C_p)/\sqrt{R_s R_p C C_p} \quad (23)$$

From (22), the following in-equality is obtained

$$(R_s C - R_p C_p)^2 + R_p^2 C^2 + 2R_p^2 C C_p > 0$$

Thus, $\zeta > 1$. It can be concluded that the buffer system is always over-damped.

It is also possible to quantify how each of the battery parameters affects $\zeta$. For example, by partial differentiating (23) with respect to $R_s$, one obtains

$$\partial \zeta / \partial R_s = (R_s C - R_p C - R_p C_p) / (4 R_s \sqrt{R_s R_p C C_p}) \qquad (24)$$

Since $C_p$ is of the order of F but $C$ is of the order of mF, the term $R_p C$ can be ignored. Thus $R_p C_p \gg R_s C$. Therefore (24) can be simplified to yield

$$\partial \zeta / \partial R_s = -R_p C_p / (4 R_s \sqrt{R_s R_p C C_p}) \qquad (25)$$

The last sensitivity function has a negative value which indicates that with an increase in $R_s$ would cause a decrease in the damping ratio of the system.

Similarly, the same approach can be applied to the study of the sensitivity of $\zeta$ to $R_p$ and $C_p$ to obtain

$$\partial \zeta / \partial R_p = (R_p C + R_p C_p - R_s C) / (4 R_p \sqrt{R_s R_p C C_p}) \qquad (26)$$

$$\partial \zeta / \partial C_p = (R_p C_p - R_s C - R_p C) / (4 C_p \sqrt{R_s R_p C C_p}) \qquad (27)$$

Both sensitivity functions have positive values. It indicates that $\zeta$ increases with $R_p$ or $C_p$.

Finally, the characteristic equation shows that there are two real negative roots

$$\begin{aligned} s_1 &= 0.5(\alpha - \beta - \sqrt{(\alpha + \beta)^2 + 4\gamma}) \\ s_2 &= 0.5(\alpha - \beta + \sqrt{(\alpha + \beta)^2 + 4\gamma}) \end{aligned} \qquad (28)$$

Upon closer examination, it is noted that $\alpha > 0$, $\beta < 0$ and $\gamma > 0$. Hence it can be concluded from (28) that $s_1 \ll s_2$ and the dynamic performance of the system is primarily dominated by the pole $s_1$.

From [7, 11], it is shown that $R_s$, $R_p$ and $C_p$ vary with the discharge current. From the data shown in [11] for example, one can obtain the values of $R_s$, $R_p$ and $C_p$ and from which the damping ratio $\zeta$ and dominant pole $s_1$ variations with the discharge current can be determined. Typical results are shown in Fig. 6. In this example, one finds that $\zeta$ reaches its maximum value at a discharge current of some 153 A before it progressively decreases as the current increases. The dominant pole changes little as the discharge current increases to 153 A but decreases greatly with further increment of the current.

In the study of buffer system operation, one should therefore select the battery condition which will result in the highest damping ratio and smallest $s_1$ as the worst operating condition. This is because during sag, a higher value of the damping ratio corresponds to a buffer system with slower response characteristic. In Fig. 6, for example, the most strenuous load ride-through condition would correspond to battery discharge current of about 153 A. Hence, by designing the buffer to provide load ride-through under this most strenuous condition, improved and smoother ride-through under other discharge current condition can be guaranteed.

IV. SIMULATION RESULTS

To verify the validity of the operation scheme proposed in this paper, a power buffer model based on 0 for a 50-Hz 415 V system was built using Matlab/Simulink platform. The parameters used in the simulations are shown in Table I.

TABLE I PARAMETERS FOR CONSTANT POWER LOAD SIMULATIONS

| $v_g$ | 415 V | $R_{s(1000A)}$ | 0.216Ω | $R_{s(153A)}$ | 0.461Ω |
|---|---|---|---|---|---|
| $R$ | 61.33 mΩ | $R_{p(1000A)}$ | 0.072Ω | $R_{p(153A)}$ | 0.288Ω |
| $L$ | 0.97 mH | $C_{p(1000A)}$ | 1.39 F | $C_{p(153A)}$ | 6.94 F |
| $C$ | 10 mF | $E_0$ | 864 V | $P_l$ | 100 kW |

Under pre-sag condition, the dc-link voltage is maintained at 859 V at the modulation index of 0.79. A constant power load of 100 kW is connected to the buffer. Incidents of 10-cycles upstream unbalanced faults are used to study the performance of the buffer system. The data of the buffer system is taken from [2] while that of the battery is from [11].

Due to space reason, only a sample of the simulation results will be included here. The case shown is pertaining to an unbalanced voltage sag which has resulted in 80% positive phase sequence plus 20% negative phase sequence components buffer terminal voltage. The control method of the power buffer is switched from constant power mode to constant impedance mode after the sag is confirmed, 2 ms from sag initiation. The simulation results are shown in Fig. 7.

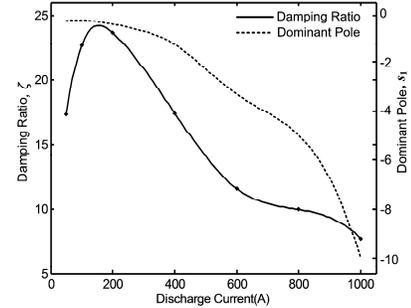

Fig. 6 Damping ratio and dominant pole versus battery discharge current.

PCC voltage waveform in Fig. 7(a) shows the occurrence of the unbalance voltage sag. Fig. 7(b) shows that the dc-link voltage (solid line) decreases soon after the sag initiation but the battery connected to the dc-link subsequently boosts the voltage and supplies the mismatch power. $V_{dc}$ is kept within desirable range despite the constant impedance control method used during the sag period. For comparison purpose, the dc-link voltage without the battery switch-in is also shown in Fig. 7(b). Without the battery, the deficiency in power between what the upstream can supply and that demanded by the load causes the dc-link voltage to decrease to zero following the sag occurrence. If this occurs, the load would be disturbed. The input current waveform of Fig. 7(c) shows that under the unbalanced disturbance condition, through the negative phase sequence system controller in the control system, the input current drawn from the supply is controlled to be essentially balanced. Since the buffer control method is under constant impedance mode over the sag period, the input current drops to 0.8 p.u. or some 160 A, which is proportional to that of the sag voltage. The buffer current overshoot observed in Fig. 7(c) is caused by the controller mode change and the switching in of the battery following the sag occurrence. The 2 ms delay assumed between the instance of the detection of fault and when the buffer control mode change and the battery switch-in has contributed toward the current transient. The transient current reaches some 25% above its pre-sag value for a very short duration. Therefore, the switching device within the buffer has to be rated to carry this transient current momentarily.

Figs. 7 (d) and (e) show the variations of input real- and reactive-powers from the upstream system into the buffer and the real power supplied by the battery, respectively. It can be seen from Fig. 7(d) that the real power supplied by the upstream is reduced to 64 kW or 0.64 p.u. of the nominal value, over the sag period, and that the input reactive power at the PCC is controlled to be zero pre- and post-sag in order to satisfy the UPF operation. From Figs. 7 (d) and (e) one also finds that the real power supplied by the upstream system during pre-fault duration meets the load demand of 100 kW and the combined power supplied by the utility and the battery during sag is maintained at 100 kW, the mismatch power being supplied by the battery energy storage system.

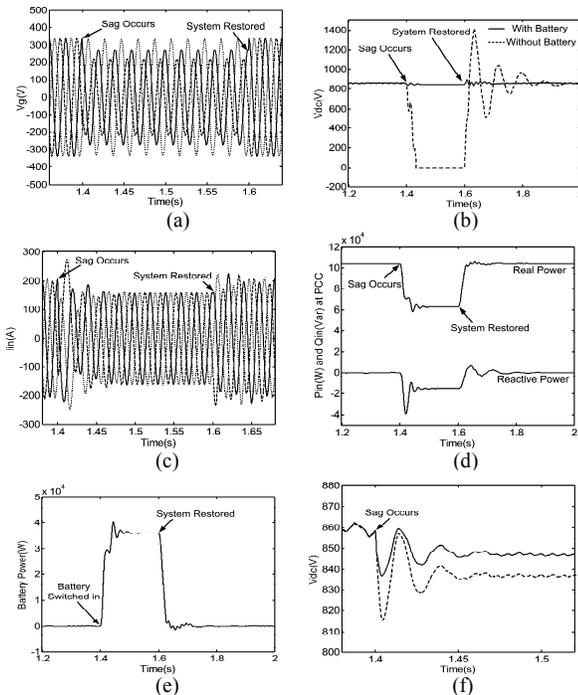

Fig. 7 Buffer-load response under an unbalanced voltage sag of 80% positive phase sequence and 20% negative phase sequence: (a) PCC voltage, (b) DC-Link voltages, with and without the battery, (c) input current, (d) input real and reactive power at PCC, (e) real power supplied by battery bank, (f) DC-Link voltage performance at discharge current of 153 A (---) and 1000 A (—).

Finally, to verify the influence of the battery parameters on the transient performance of buffer system, two sets of battery parameters corresponding to different discharge current conditions have been used in the simulation. In this example, the same battery used in obtaining Fig. 6 has been assumed herewith. Fig. 7(f) compare the $V_{dc}$ response corresponding to discharge current of 1000 A and that of 153 A. One notes that the response of the latter case, as depicted with dotted line, is less well-damped than that of the higher discharge current. This verifies the finding described in Sub-section III.C.

V. CONCLUSIONS

A new scheme for the design and operation of power buffer is proposed. Under normal network condition, the buffer is to operate under constant power mode. When voltage sag occurs, the buffer is switched into constant impedance mode, so as to mitigate the negative impact of the constant load on voltage stability of the system. The mismatch power between the upstream source and that demanded by the load during the sag is supplied by the battery bank connected to the buffer dc-link. Voltage stability of the power system is therefore enhanced. Furthermore, analytical expressions governing the load ride-through capability and the battery state-of-discharge have been derived. Through an analysis of the buffer-battery transient response characteristic, the most onerous condition under which load ride-through can be achieved has been identified. The steady state and transient performances of the proposed buffer system are verified through simulation. The results have demonstrated the effectiveness of the proposed scheme in mitigating voltage instability due to disturbances occurring in the upstream supply system.